\documentclass[12pt]{article} %
\usepackage[english]{babel}
\usepackage{ae,subfigure,parskip}
\usepackage{amsmath,amssymb,amstext}
\usepackage{graphicx,cite,a4wide}
\usepackage{xspace,subfigure}
\usepackage{epsfig}
\usepackage{booktabs}
\usepackage{color}
\usepackage{todonotes}
\usepackage{verbatim}
\usepackage{nicefrac}
\usepackage{hyperref}
\usepackage[version=3]{mhchem}

\newcommand{\tsp}[1]{\textsuperscript{#1}}

\graphicspath{{pics/}}

\begin{document}

\thispagestyle{empty}

\mbox{}

\vspace{1cm}

\begin{center}

\LARGE\bf Global optimization of parameters in the reactive force field
          ReaxFF for SiOH\\[3cm]
   \Large\it H.~R.~Larsson$^\dagger$,
             A.~C.~T.~van Duin$^\ddagger$
             and B.~Hartke$^{\dagger\ast}$
             \\ [1cm]
   \large\rm $\dagger$ Institut f\"{u}r Physikalische Chemie,\\
             Christian-Albrechts-University,\\
             Olshausenstr.~40,\\
             24098 Kiel,\\
             Germany\\[5mm]
             $\ddagger$ Department of Mechanical and Nuclear Engineering,\\
             The Pennsylvania State University,\\
             University Park,\\
             Pennsylvania 16802\\
             USA\\[5mm]
             $\ast$ corresponding author, hartke@pctc.uni-kiel.de\\[3cm]
    \large\it J.~Comput.~Chem. \textbf{2013}, \textit{34}, 2178-2189, \url{https://doi.org/10.1002/jcc.23382}.\\
\end{center}

\newpage

\thispagestyle{empty}

\mbox{}

\vspace{1cm}

\section*{Abstract}

We have used unbiased global optimization to fit a reactive force field to a
given set of reference data. Specifically, we have employed genetic algorithms
(GA) to fit ReaxFF to SiOH data, using an in-house GA code that is
parallelized across reference data items via the message-passing interface
(MPI). Details of GA tuning turn out to be far less important for global
optimization efficiency than using suitable ranges within which the parameters
are varied. To establish these ranges, either prior knowledge can be used or
successive stages of GA optimizations, each building upon the best
parameter vectors and ranges found in the previous stage. We finally arrive at 
optimized force fields with smaller error measures than those published 
previously. Hence, this optimization approach will contribute to converting 
force-field fitting from a specialist task to an everyday commodity, even for 
the more difficult case of reactive force fields. 

\vspace{2cm}

{\bf keywords:} global optimization, force-field fitting, genetic algorithms,
evolutionary algorithms, reactive force fields

\newpage

\setcounter{page}{1}

\section{Introduction}
\label{section:intro}

Chemistry deals with reactions between molecules. The inter- and
intramolecular forces governing these reactions result from the electrostatic
interactions between the atomic nuclei and the electrons, and from the
fundamentally quantum-mechanical nature of the latter. Frequently, it is
possible to obviate the need for an explicitly quantum-mechanical treatment by
modeling interatomic forces with empirical potentials or force fields. This
incurs savings in computer time of about six orders of magnitude, which makes
it possible to simulate large-scale systems with billions of atoms
\cite{billion} or somewhat smaller systems for very long times, currently up
to the millisecond regime \cite{Shaw,Pande} for protein folding, i.e.,
covering 10\tsp{12} time steps. In biochemistry simulations, non-reactive
force-fields are employed almost exclusively. Nevertheless, various reliable
reactive force-fields are also in common use in other areas of
chemistry.\cite{Brenner,Miller,Warshel1,Truhlar,Warshel2,Warshel3} 
Several of them are specific for certain 
systems or atoms.\cite{Johnston,BoldingAndersen,Singer,Jones} With
ReaxFF a rather general reactive force-field has become available in the last
decade\cite{ReaxFF_CH}. After the first parametrization for hydrocarbons
\cite{ReaxFF_CH}, ReaxFF parameters have been produced for various other
systems, ranging from inorganic materials \cite{ReaxFF_HBN} via various
organo-catalytic transition-metal systems \cite{ReaxFF_CoNiCuCH,ReaxFF_VCHO}
to pure metals \cite{ReaxFF_Au}. Hence, ReaxFF is not limited to certain
classes of compounds, 
and it can handle both periodic and molecular systems. The character of the 
force field implementation is kept modular, i.e., there should exist only one 
set of parameters for each atom which can be used in combination with parameter 
sets of other atoms. 
Despite the greater computational cost of a
more complicated force-field like ReaxFF, it was also shown that it is
amenable to large-scale simulations \cite{ReaxFF_large}.

However, there is no known way to extract analytically accurate force fields
from the molecular Schr\"{o}dinger equation. Additionally, the accuracy
requirements of computational chemistry are very strict. Therefore, the
standard approach is to fit force fields to reference data, which can be an
arbitrary mixture of experimental data and \textit{ab-initio} quantum-chemical
calculation results.

Since fitting empirical force fields to reference data via variation of the
force-field parameters is a high-dimensional and non-separable optimization
problem with multiple minima, standard gradient-based local search techniques
are of little help. Deterministic global optimization techniques
\cite{Neumaier1,Neumaier2} are available but face severe practical
difficulties in high-dimensional search spaces and for computationally
expensive objective functions. Unfortunately, both characteristics are typical
for the present problem. Viable alternatives are non-deterministic search
heuristics, for example, genetic algorithms (GA)
\cite{Holland,Goldberg}, or more generally evolutionary algorithms (EA). 
They offer comparatively very fast access to very
good solutions, at the expense of failing to guarantee global
convergence. This is ideal for the present task, since there is no need to
find the global minimum, as long as a solution actually found is not very much
worse. 
{%
Of course, also other non-deterministic global search methods have been
applied to the force-field fitting task, for example simulated annealing
\cite{SAparamfit1,SAparamfit2,SAparamfit3} or artificial neural networks
\cite{ANNparamfit}. A complete overview is impossible here.}

Hence it is no surprise that GAs have been frequently used for force-field
fitting, since many years. 
{%
Similar GA approaches also are in use in many other areas, for
example in determining the cluster expansion
\cite{clusterexpGA1,clusterexpGA2} for predicting properties of 
alloys. Not surprisingly, these GA applications share common characteristics.}
Almost all of the GA applications to force-field fitting, however, were
directed at non-reactive force-fields, reflecting their prevalence. We are not
attempting to give a comprehensive overview here but restrict ourselves to a
few characteristic examples.

The study that claims to be the first one of this kind was presented in 1998
by Hunger \textit{et al.} \cite{Huttner1}. There, GA-optimization was used to
generate new MM2 parameters, specifically for tripod-Mo(CO)$_3$ compounds,
with experimental structural data as reference. To save time in evaluating the
force-field data, in an extension of this work \cite{Huttner2} a neural
network was trained on the results of several previous GA generations, thus
saving several orders of magnitude in computational time.

Subsequently, several studies employed GA-methods to produce new or better
parameters for various chemical species in standard force-fields like MM
\cite{Cundari,Courcot}, MM3 \cite{Strassner,Schmid} and AMBER \cite{Kollman}
as well as in simpler special-purpose force-fields \cite{Globus,vdVegt}. In the
latter category, there are also early works by one of the present authors
\cite{gaga,si_n,water567}, which partially pre-date the publications by Hunger
\textit{et al.} mentioned above.  In most of these cases, DFT data were used
as reference, but obviously also any other level of theory can be used,
including 
CCSD(T) \cite{lj_mixed}. It was also investigated \cite{Gelb} how to adjust
technical details of the GA itself (mutation, crossover, population size,
selection) for best performance in such tasks.

In all these GA applications, the immediate difficulty arises that the
reference data set contains a broad variety of items of different nature and
with different impact on the parameter optimization. Usually, each item
receives a specific weight, and these weights are then included into a sum of
squared deviations between force-field and reference items, as a single
objective, which in turn is minimized. Obviously, there is no unique solution
to this task, in the sense that different weights will produce different best
solutions (even if there were only one single best solution for each set of
weights, and assuming that this solution can be found with certainty). 

This problem arises for very many optimizations of practical interest, and
multiobjective optimization is frequently advertized as a solution --- also
for the task of force-field fitting. For example, Handley and Deeth
\cite{Deeth} used multiobjective-EA-optimization of a special ligand-field
molecular mechanics (LFMM) force-field for spin-crossover iron-(II)-amine
complexes. %
Indeed, such methods are able to simultaneously deliver a balanced set of
solutions (the Pareto front or set) in which each member excels in one
particular sub-objective at the cost of lesser performance in all others. In
the end, however, it is still up to the user to decide which of these
solutions to actually adopt as the single best one for a specific
purpose. Hence, it is not immediately clear that multiobjective approaches are
superior to executing single-objective methods several times with different
sets of weights.

In all applications mentioned so far, a fixed functional form is pre-assumed
for the force field. It typically is not able to adopt all shapes that are
mathematically or even physically possible, i.e., this choice constitutes a
bias and puts a lower bound on the force-field deviation that is greater than
zero. Theoretically, it is indeed possible to search for globally optimal
functional forms for interatomic potentials in a less biased fashion, while
simultaneously fitting parameter values. This was realized
\cite{Thompson,Thompson2} using a combination of genetic programming,
Monte-Carlo sampling and parallel tempering. These attempts, however, while
impressive, have not yet advanced beyond the proof-of-principle stage.

Compared to the huge amount of work on GA-optimization of non-reactive
force-fields, parts of which was just described, almost no work of this kind
has been done so far on reactive force-fields. Only recently, Angibaud
\textit{et al.} \cite{Briquet} performed a GA-optimization of ten
silicon-silicon interaction parameters in a charge-transfer reactive (CTR)
force-field for silica \cite{CTR}, without local search. A later extension to
silicon-carbon and carbon-carbon parameters \cite{Philipp} also employed
GA-optimization. %
The only other study of this kind \cite{Pahari} 
that has come to our attention, and the only one that has targeted ReaxFF,
will be discussed in some detail below.

This state of affairs is somewhat surprising, considering that already in 1994
(again pre-dating the work by Hunger \textit{et al.} \cite{Huttner1}) Rossi
and Truhlar \cite{RossiTruhlar} employed a GA to generate new semiempirical
NDDO parameters, specifically for the reaction Cl + CH$_4$, in support of the
idea of specific reaction parameters (SRP) \cite{SRP}. Clearly, from the
viewpoints of parameter optimization and of the later use of the fitted
function, there is little difference between Rossi and Truhlar's work on the
one hand and reactive force-field fitting on the other hand. In fact, the
latter is even simpler, for several reasons: Function evaluation is faster and
the influence of the parameters on the final energy is more direct. Therefore,
the present article helps to fill an obvious gap, by employing GA methods
for parameter optimization in a reactive force-field of the ReaxFF
type. Clearly, employing a global optimization algorithm is even more
important for a reactive force-field than for a non-reactive one, since the
former kind typically has more parameters per atom than the latter.

The more specific purpose of the present contribution is twofold. One aim is
to demonstrate the practical feasibility of GA optimization of ReaxFF
parameters, for a given test case. The given items consist of the reference
data set, a set of ReaxFF parameters to be optimized, and allowed ranges for
all of these parameters. We have investigated how to ``tune'' the GA for this
purpose. It has been criticized \cite{Goedecker} that GAs need too much
adjustment, compared to other non-deterministic global optimization
algorithms, implying that the latter are more robust and hence simpler to
use. Our past experience as well as our present findings do not corroborate
this: As shown in section \ref{section:GAopt.tune}, the influence of GA tuning on
GA performance is much smaller than anticipated. Proper adjustment of other
aspects of the optimization is far more important, and these adjustments would
be needed in exactly the same way for other global optimization algorithms.

A second aim is to explore if previously obtained ReaxFF fitting results can
be reached or even improved upon. Most of the presently available ReaxFF
parameter sets were obtained using a successive one-parameter parabolic
extrapolation technique \cite{soppe}, here abbreviated as \textsc{soppe}
(which is not strictly a local search algorithm but definitely also does not
pretend to do global search), in conjunction with multiple restarts and a
large amount of human experience. With the present contribution, we indeed
improve upon previous ReaxFF parameter sets, hence demonstrating
that at least parts of this human fitting experience can be substituted by
suitable series of GA runs. This lessens the burden of force-field
fitting and makes it more accessible to the non-expert.

In real-life applications, much of what is prior information here has to
be established first: reference data, parameters to be optimized, and their
ranges. In preliminary studies \cite{KomPlasTech2013}, we have already shown
that GA parameter fitting also is applicable, efficient, and successful in
{%
situations in which this information is missing}, 
for the example of azobenzene in its electronic ground
state. Ongoing work is also establishing this for the electronically excited
states of the same molecule. In these applications, the techniques
established here for SiOH are re-used largely unchanged,
{%
after obtaining
reference data, parameters to be optimized, and their
ranges in a simple trial-and-error fashion}.

In the final phase of our work for the present project, Pahari and Chaturvedi
\cite{Pahari} published a paper in which they also proposed the idea of using
GA optimization to fit ReaxFF parameters to reference data. In their study,
however, the actual GA optimization is applied only at the very end. The focus
is almost exclusively on preparatory steps prior to the GA, namely on
determining the smallest set of ReaxFF parameters that have the largest
influence on the force-field error, using sensitivity tests and cross-correlation
information. Additionally, for the whole study, they have restricted the
reference data set to a single molecule (nitromethane) and four of its
decomposition products, for which exclusively 
geometry data of locally
optimized structures are employed as references. Furthermore, they initiate
their GA series from a single, pre-determined trial vector, and do not employ
local optimization. In contrast, in our studies presented here, the focus is
on the GA part. We start our initial set of GA calculations with randomized
parameter vectors, and we employ the option to hybridize the GA with local
search, which turns out to be crucial in the ``endgame''. Instead of
attempting to narrow down the initial set of parameters to be optimized, we
stick to it or even attempt enlargements. Last but not least, we employ a
large and diverse set of reference data, containing many different species and
also very different properties (in addition to some geometry data, also
partial charges, crystal cell parameters, and many relative energies of
different species). Nevertheless, in the real-life situations discussed in the
previous paragraph, and if it is possible to quickly calculate the force-field
error of very many trial vectors, the techniques presented by Pahari and
Chaturvedi definitely are very important and useful.

The remainder of this article is organized as follows: In section
\ref{section:methods}, we provide details for all general approaches employed
in this work. Section \ref{section:GAopt} focuses on how to ``tune'' the GA
to the optimization case at hand. The actual application of our techniques to
the SiOH case is described in section \ref{section:SiOH}, which is split in a
first subsection \ref{subsection:SiOH.production} describing procedural
details, and a second subsection \ref{subsection:SiOH.results} where we show
and discuss the results. The final section \ref{section:conclusions} provides
a summary and an outlook to future work.

\section{General methods and computational techniques}
\label{section:methods}

\subsection{{Algorithmic details}}
\label{section:methods.algo}

In the program design phase for the present project, we anticipated that
reference-data-level parallelization would turn out to be a decisive asset to
make large-scale global optimization feasible within acceptable real-time.
Another design goal was to stay as close as possible to the original ReaxFF
implementation. These two goals could only be met by a tight, deep-level
interfacing of a GA code and of the ReaxFF code. Therefore, we decided not to
use the universal GA program suite \textsc{ogolem} \cite{ogolem}, the
development of which was started in the Hartke group. Instead, we wrote a new
GA implementation which directly addressed suitable pieces of almost
unmodified ReaxFF code.

The GA part of the code is based on a standard GA, as described by Goldberg
\cite{Goldberg}, with the following exceptions: There is no binary encoding of
parameter values; instead, the genetic string simply is an array of reals, for
all ReaxFF parameter values to be optimized. A single-point crossover is
employed; the crossover point is restricted to the parameter boundaries and is
determined via evenly weighted random numbers. Mutation of single parameters
is realized via normal-distributed deviations from the parent parameter
value. The number of parameters to be mutated is an additional external
tunable, the mutation amount. Mutation (to a given amount) is always done if
crossover does not happen (otherwise, mating would degenerate into a simple
but unproductive parent copy). Most importantly, the concept of generations is
abandoned; instead, a steady-state or pool model is used \cite{ogolem,Bandow}:
From the population of trial solutions, pairs of individuals are extracted
quasi-continuously (or, in one of the parallelization options discussed below,
in an embarrassingly parallel fashion), selected by suitably designed fitness
functions. GA operations are performed upon the selected individuals, and the
resulting children are staged for re-entering into the population according to
their fitness. If a child is fitter than the currently worst individual in the
pool, it replaces the latter; otherwise the child is discarded. To maintain a
controllable level of diversity in the pool, diversity measures are introduced
via two additional external tunables. One of them is a percentage threshold
value that determines if two given values for the same parameter are taken as
identical or as different. The second one determines the fraction of all
parameters that have to differ (according to the previous measure) such that
two individuals are regarded as different. During re-placement of children
into the pool, it is strictly avoided to arrive at two or more pool
individuals that are ``identical'' according to these criteria. If two
``identical'' individuals occur, the better one survives, the other one is
discarded. After a number of global optimization steps input by the user, each
new child with a fitness below a user-supplied threshold value is subjected to
a local optimization (discussed in the next paragraph). Following established
GA usage in the field of global cluster structure optimization
\cite{AngewReview,WIREsReview}, we first started our GA production runs with
locally optimizing each and every new child, from the very beginning. However,
in contrast to cluster structure optimization, analytical first derivatives
are absent, which renders local optimization expensive. In the present GA
application, this increased expense outweighs the increased optimization
power. Hence, we changed our approach to performing the first 98--99\% global
optimization steps without local optimization in parameter space (as ``single
points''), switching on local optimization only in the final 1--2\% of the
steps. At the switching point, the whole pool is locally optimized, otherwise
selection would render the unoptimized part of the pool obsolete.

\subsection{{Local optimization}}
\label{section:methods.local}

Since calculation of ReaxFF energies involves an iterative scheme to determine
partial charges, and since items in the training set can additionally involve
an iterative geometry optimization, there are no simple derivatives of ReaxFF
energies with respect to force-field parameters. Therefore, we have tested and
employed four different local optimization algorithms without derivative
information: an older constraint-free algorithm by Powell, as published in the
``Numerical Recipes'' \cite{NumRec}; 
\textsc{newuoa}, a
constraint-free algorithm by Powell \cite{newuoa}; \textsc{bobyqa}
\cite{bobyqa}, a newer variant of \textsc{newuoa} with constraints;
and the original \textsc{soppe}
strategy \cite{soppe} of ReaxFF (which also obeys
constraints on the parameter variations). Generally, \textsc{bobyqa} and
\textsc{newuoa} were found to be vastly superior to the older Powell
algorithm. In fact, also in other applications tested in the Hartke group,
\textsc{bobyqa} displayed superior efficiency. It also outperforms
\textsc{soppe}; nevertheless, the latter was also employed in the final
phases, since due to its rather different iteration directions it may still
make progress when \textsc{bobyqa} stagnates. For practical usage of both
\textsc{bobyqa} and \textsc{newuoa}, it is recommended that the ranges of all
coordinates of the search space should have similar extent. This is decidedly
not the case for the ReaxFF parameters to be optimized here. Hence, for use
inside these local optimization routines, all ReaxFF parameters were
temporarily re-scaled to the interval $[0,1]$.

\subsection{{Parallelization}}
\label{section:methods.parallel}

There are at least three obvious levels of parallelization possible in our
setting: (1) across GA individuals, (2) across reference data set items, and
(3) within ReaxFF energy evaluations. The latter is the standard option of
calculating interparticle contributions only if they are within distance
cutoffs, which breaks the calculation down into quasi-independent,
neighborhood-based chunks. This is utilized in the ReaxFF implementation
within ADF \cite{ADF_ReaxFF}
as well as in many other force-field-based MD
program packages,
and will not be discussed further here. The first two options both offer
embarrassing parallelism and hence excellent scaling. Option (1) is
conceptually trivial within the pool variant of the GA and leads to complete
elimination of serial bottlenecks, as the Hartke group has shown earlier
\cite{Bandow}. To simplify coding for the present project, this option was not
employed here. Instead, we have focused on option (2), as described in the
next paragraph.

Using explicit messaging via the message-passing interface (MPI), we have set
up a straightforward master-slave model. For every given GA individual (which
simply is a single set of ReaxFF parameter values), the master hands out
reference data items to the slaves which produce the corresponding ReaxFF
comparison values for them. The advantage of the master-slave paradigm is that
only the master steps through all GA-related code parts (which are not
compute-intensive, compared to the ReaxFF calculations), while the slaves
enter the reference item treatment section once at the beginning of program
execution and then stay trapped there until the end of the program. This
ensures a clean program code, despite the very basic MPI
parallelization. Handing out reference data items is realized via a simple
loop that ensures automatic load balancing \cite{usingMPI}. In our initial
implementation, only single reference data items are handed out. Since the
computational time needed for many of them is rather short (cf.\ section
\ref{section:GAopt.reduce}), increasing the number of slaves beyond a certain number
will diminish overall speedup due to communication overload at the
master. Nevertheless, we can achieve reasonable speedups with a few dozen
slave processes.

\subsection{{Test cases}}
\label{section:methods.testcases}

We have {started to perform} global GA optimization 
of ReaxFF parameters for three test
cases: Co, SiOH and Gly. For each of these cases, the van Duin group
prescribed a reference data set, a set of ReaxFF parameters to be optimized,
and lower/upper limits within which the ReaxFF parameters were allowed to be
varied. With some exceptions described in sections \ref{section:GAopt.ranges} and
\ref{subsection:SiOH.production}, we have not changed these pre-settings.

In all of these cases, we have used the same general measure for deviations
between the force-field data and the reference data, which we will abbreviate
\textit{error sum} in the following:
\begin{equation}
\label{eq:error_sum}
  \text{error sum} = \sum\limits_{i=1}^n
  \left(\frac{x_{i,\text{calc}}-x_{i,\text{ref}}}{\sigma_i}\right)^2.
\end{equation}
$x_{i,\text{ref}}$ is the value of a reference data item, and
$x_{i,\text{calc}}$ is the corresponding value as calculated from
ReaxFF. 
{
$\sigma_i$ serves several purposes: Formally, it is a standard deviation
and a way to make each summation term dimensionless. Practically more
important is that the user is free to introduce arbitrary numerical values for
it, which makes it possible to tune the relative weight of each item as needed.
Therefore, individual error sum values are rather meaningless, they could be
scaled to any number larger than zero. Additionally, due to the different
number and nature of reference items, they 
are not comparable anyway between different cases. Hence, their only purpose
is to gauge relative optimization progress within one given case.}

The Co test case is based upon recent work in the van Duin group \cite{Co},
where a ReaxFF parametrization for cobalt was developed and shown to yield
good descriptions of the energetics and properties of various crystal phases,
amorphous configurations, clusters, vacancies and surfaces. 12 ReaxFF
parameters were to be varied, and the reference data set consisted of 147
items. Without using further information, {and with minimal
computational effort,} the best error sum of 
1443.7
achieved previously by the van Duin group could be improved in our GA
optimizations to 1439.8; no further progress seems to be
possible. {Hence,} compared to the SiOH case treated in the
remainder of this article, {parameter optimization tasks with a
  complexity of this Co case pose no challenge to our approach}.

{
The Gly test case is work in progress. Its details and current status will be
mentioned briefly in the conclusion section \ref{section:conclusions.outlook}.}

The remainder of this article is focused on the SiOH test case,
{which poses a surmountable challenge to our approach. This test
  case} is based 
upon earlier work of the van Duin group \cite{SiOH,SiOH2}, where a ReaxFF
parametrization for materials involving silicon and silicon oxides was
developed. In its original form, it contains 67 parameters to be
optimized, 304 reference structures (ranging from simple molecules like
silanes to periodic solid-state structures like quartz) and a set of 309 
reference properties for them (the training set, consisting of geometry data,
partial charges, crystal cell parameters, and relative energies). 

\subsection{{Factors influencing GA performance}}
\label{section:methods.GAperf}

In real-life applications of the present global parameter optimization scheme,
there is less information available in the beginning. Namely, it is as yet
(largely) unknown which reference data items should be used, which ReaxFF
parameters should be varied and within which limits these variations should
happen. It will be a recurring theme of the next two sections that these items
have a decisive influence not only on the final force-field quality but also
on the overall efficiency of the GA optimization. To take the example of
parameter variation ranges, it is clear that varying the ranges directly
influences the size of the search space volume to be covered by the
optimization. Trivially, a search in a smaller space is much more efficient
than a search in a larger one. Therefore, starting from good guesses for the
parameter ranges is vital for obtaining good results within short real
times. Conversely, allowing the full range $[-\infty,+\infty]$ for all
parameters would be a bad idea (disregarding the possibility that this may
lead to unphysical characteristics of various force-field terms, for many
parameter value realizations). Fortunately, it turns out that a
straightforward iteration of the whole GA scheme allows for narrowing down the
variation ranges even in the absence of prior information on them. 
{This
range-fitting procedure will be explained in the next section
\ref{section:GAopt.ranges} and applied in section \ref{section:SiOH}.}

As already mentioned in the introductory section \ref{section:intro}, another
decisive lever influencing the global optimization is to change the number of
parameters to be optimized, or their selection. How to do this systematically
and successfully, has already been demonstrated by other authors
\cite{Pahari}. In the present study, we employ{ed} this option to generate
smaller test scenarios, to make GA tuning faster (cf.~section
\ref{section:GAopt.reduce}), and to attempt a significant enlargement of the set of
parameters to be optimized, to further improve the error sum (cf.~section
\ref{subsection:SiOH.production}). The central idea of this study, however, was
to take the set of parameters to be optimized as given and immutable, even if
it presents us with a large search space, and to demonstrate feasibility of GA
optimization within this given search space.

\section{Optimizing GA performance}
\label{section:GAopt}

\subsection{{Reducing test set size for GA tuning}}
\label{section:GAopt.reduce}

As quantitative measurement of the quality of the force field, the error
sum of squares {was} used, defined as already shown above in
eq.~\ref{eq:error_sum}.  
The values of the reference data items, $x_{i,\text{ref}}$, were
obtained from B3LYP/6-31G$^{\ast\ast}$ calculations\cite{SiOH}. 
Various data items are possible, for example
bond lengths, partial charges or relative energies, all these either for fixed
input geometries or evaluated after a geometry optimization. The comparison
values $x_{i,\text{calc}}$ are calculated in the same fashion (i.e., possibly
preceded by a geometry optimization) from
ReaxFF. 

To obtain good performance of the GA, its parameters (some of which were
mentioned in section \ref{section:methods}) needed to be optimized. The SiOH
training set 
consists of 304 individual computation jobs for comparison with the reference
data. This turned out to be too much for a quick but thorough investigation of
the behavior of the GA parameters and needed
to be reduced for this purpose. Hence, the needed time for each of the 304
computations for 247 randomly created force field parameter vectors was
retrieved and compared. Most of the jobs needed only milliseconds
but some of them, e.g.\ geometry optimizations of periodic
structures, up to two minutes. Removal of all jobs that needed more
than 0.5 seconds (90 jobs altogether) yielded a time saving of about 95\%.  
However, the test cases for the GA parameter optimization should be small,
i.e.\ not time consuming, but also representative. Therefore, an additional
test case was obtained by removing less jobs and by ensuring that
the reduced training set still consisted of jobs of different types. 43 jobs were
removed which yielded a still useful time saving of 87\%. 

To get insights into the behavior of the GA by changing the dimension of the
optimization problem, the set of 67 to be optimized parameters was reduced to
40. With these reductions, four test cases were obtained, each of them with 214
or 261 jobs in the training set and 40 or 67 parameters to be
optimized. Table \ref{tab:settings} sums this up. 

\begin{table}[htbp]
\centering
\caption{Scheme of the test settings.}
\label{tab:settings}
\begin{tabular}{lccp{5cm}}
 \toprule
  case & parameters to be optimized & training-set-jobs\\ \midrule
  bo & 67 & 261 \\
  bp & 40 & 261  \\
  so & 67 & 214  \\
  sp & 40 & 214  \\
  original & 67 & 304  \\
 \bottomrule
\end{tabular}
\end{table}

\subsection{{GA tuning}}
\label{section:GAopt.tune}

As in all indeterministic global optimization algorithms of this kind, there
is no true convergence criterion. Further progress does diminish
progressively, but ultimately the user has to decide when to stop, and this
depends on the problem at hand. After some tests containing 20 different GA
runs with a randomly created starting population and a maximum iteration number 
of 1000 with each of the four test cases and three runs with a maximum iteration
number of 10000  with the case ``so'' (see Table \ref{tab:settings}), a total
iteration number of 800 was found to be sufficient to reach 
the region where progress ha{d} essentially leveled off. 

{
Then, 19 tests were performed, with a systematic change of one GA-parameter. 
Each of these tests consisted of 10 GA runs for each of the four test cases,
resulting in a total of 760 single GA test runs. Afterwards,} 17
additional tests were performed with a systematic change of two parameters at
once, to check for 
correlations between parameters. The analyzed parameters were the initial
pool size, the overall pool size (which can be different from the initial pool
size, i.e., the pool is initially allowed to grow), the crossover rate, the
mutation amount and the two diversity threshold parameters 
mentioned in section \ref{section:methods.algo}. 

\begin{figure}[htbp]
  \centering
  \includegraphics[width=0.8\textwidth]{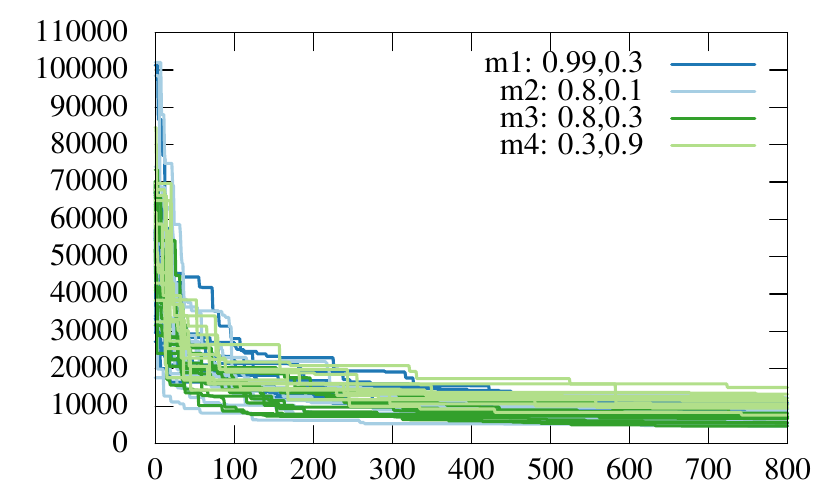}
  \begin{minipage}{0.8\textwidth}
  \caption{\label{figure:gaopt_mut} Selection of test runs
    with varying crossover probability and mutation amount. For each 
    iteration in 10 runs of each kind, the error sum of the currently
    best individual is shown. The internal identifier is given for each run,
    as well as the crossover probability and the mutation amount. The color
    scheme for this figure and for later ones was taken from
    \cite{colorbrewer}.}  
  \end{minipage}
\end{figure}

\begin{figure}[htbp]
  \centering
  \includegraphics[width=0.8\textwidth]{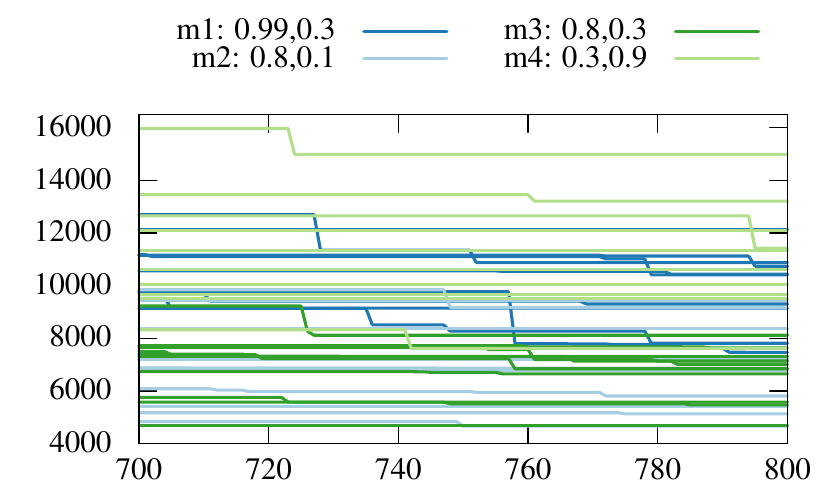}
  \begin{minipage}{0.8\textwidth}
  \caption{\label{figure:gaopt_mut_zoom} Enlarged final portion of
    fig.~\ref{figure:gaopt_mut}, for global iterations 700 through 800,
    to make the individual traces better visible.}
  \end{minipage}
\end{figure}

To give an example, Figs.~\ref{figure:gaopt_mut} and
\ref{figure:gaopt_mut_zoom} show the results for a semi-random selection of
some test runs of setting ``bo'' with varying crossover probability and
mutation amount.  In these kinds of figures, we plot the error sum of the 
currently best 
individual vs.\ the global iteration counter. The curves exhibit a
staircase-like shape since the best individual is not improved at every step
but is guaranteed to survive if it is not improved upon, via the automatic
elitism inherent in the pool strategy.

Surprisingly, all runs shown perform{ed} similarly, although the crossover 
probability and the mutation amount changed drastically for these runs: From 
0.99 to 0.3 and from 0.3 to 0.9, respectively. The same finding {was} obtained
from our other GA parameter test runs. 
These results are counter-intuitive if one is 
familiar with the literature where the values of the GA parameters and the
fine details of GA operator design are 
discussed extensively.\cite{weiseit,Gelb,Davis1,Davis2,Gecco}
While the change in performance by varying each GA parameter was smaller than
expected, 
it was still there. Upon first glance at Fig.~\ref{figure:gaopt_mut}, one
clearly  
sees that case ``m4'' (excessive mutation) perform{ed} slightly worse than the
others, i.e.\ there {was} less decrease in the  
error sum of the best individual during the iterations than in 
other runs. Likewise, it is discernible that case ``m1'' (excessive crossover) 
perform{ed} also worse than the remaining two
cases, especially in the beginning, until about 500 iterations. The limited 
performance of these two cases is a sign that both mutation and crossover
indeed work{ed} as intended, i.e., they {were} not superfluous decoration.

It is quite hard to decide whether case ``m2'' or ``m3'' { were} better. 
Upon closer inspection of Fig.~\ref{figure:gaopt_mut_zoom}, one sees that some
runs of case 
``m2'' arrive{d} at better results than the runs of case ``m3''. However,
there { were} also runs that perform{ed} worse and end{ed} up at error sums between 
10,000 and 8,000. It is not clear whether these findings should be attributed
to the 
limited number of runs for each case (10) or indeed {were} an outcome of
the changed mutation amount.

Investigations like these were performed for all 36$\times$4 tests mentioned
above. Analysis of the results
was assisted by an exponential fitting of all 10 runs for each test. 
Overall, the following parameter combinations showed the best performance for 
all 
four test settings mentioned in Table~\ref{tab:settings}: A population size of
about 30, a crossover probability of 0.8, a mutation amount of 0.3, a value of
0.01 for 
the threshold for taking scaled parameters as identical and 80\% as the amount 
of parameters that has to differ for two individuals to be counted as
different (see section 
\ref{section:methods.algo} for further explanations). 

The choice of population size was not as clear as the crossover/mutation
choice illustrated above. Runs with population sizes of 27 
and 100 gave similar results for the best individuals. For a
larger population, the population-filling in the beginning of an 
optimization (i.e.\ creating individuals and calculating their error sum) 
needs
more time but then the diversity is larger. During the production phase 
for the results shown in section~\ref{subsection:SiOH.production}, it became 
clear that
the convergence { was} worse for large numbers of iterations and a predetermined 
starting population and it was shown that the improved diversity with this 
population size was not really needed.

As it was shown above, it makes no 
sense to discuss whether e.g.\ the crossover probability should be 0.8 or 0.85, 
as the difference in performance is normally negligible. However, {
the parameter set quoted two paragraphs earlier (30, 0.8, 0.3, 0.01, 80\%)
as the best one of our GA tuning}
proved itself during production runs with the  
non-reduced training set mentioned in section \ref{subsection:SiOH.production}. 
{It did show}
better performance than an \emph{ad hoc} configured setting with the values 27, 
0.7, 0.2, 0.1 and 80\%. It should be mentioned that an 
adjustment e.g.\ of the diversity parameters can be worthwhile during runs in 
later production phases with a pre-arranged starting population with good error 
sums, to avoid premature convergence. 

Furthermore, four different modes of selecting pairs of individuals for mating
from the error-sum-ordered pool were tested, using
the case ``bo''.
One selection mode was to have two exponential fitness functions with
different slope, reflecting the combined need for exploitation (steep function
strongly favoring the best) and for exploration (less steep function also
giving worse individuals a chance). Another ansatz was to use step
functions.\cite{weiseit,hoffmann} Hence, the second selection mode was to use a 
Heaviside function for the first
individual, with 
an equal probability for the first half of the population, and to select the
second individual with an equal probability across the whole population. The
third selection mode
used the mentioned Heaviside function for both selections. The fourth and
last mode employed a 8-step function for both selections. The 
individuals in the first eighth of the ordered population had a probability of
\nicefrac{8}{36}, the individuals in the second eighth had a probability of
\nicefrac{7}{36} and so on. Surprisingly, all four ans\"atze showed similar
behavior, despite their varieties. This { was} unexpected because the selection
function is said to have a decisive influence on the behavior of the global
optimization.\cite{weiseit,hoffmann,Davis1,Davis2,Gecco}
Within the small performance differences found, the first selection mode
mentioned above produced somewhat better results, followed by the third one.

\subsection{{Fitting procedure for parameter ranges}}
\label{section:GAopt.ranges}

\begin{figure}[htbp]
  \centering
  \includegraphics[width=0.8\textwidth]{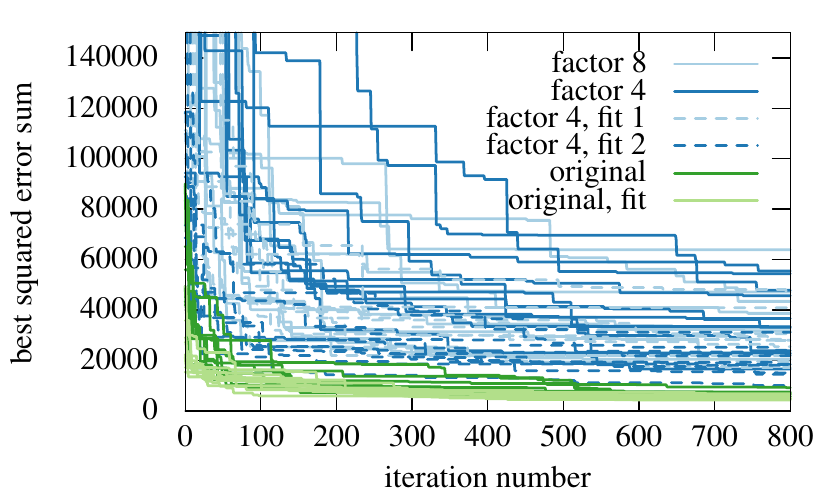}
  \begin{minipage}{0.8\textwidth}
  \caption{\label{figure:range_fitting}Results of the fitting procedure for
    parameter ranges, as explained in the text.}
  \end{minipage}
\end{figure}

Another sensitive lever are the variation ranges prescribed for each force
field parameter, since these
directly influence the size of the search space. The ranges supplied by the van
Duin group were not based on investigations by a GA, hence it was conceivable
that the GA could yield better results with a larger search space. 
{Additionally, in real-life applications of our GA parameter
optimization, these ranges usually are not available initially and have to be
established prior to or during the optimization.}
Therefore,
test runs were started using the originally given ranges with enlargement
factors of 4 and 8, respectively, with the test cases ``so'' and ``bo''. Not
surprisingly, some of these drastic enlargements {led} to unphysical behavior
of the corresponding force-field terms, which in turn induced program crashes.
Therefore, the ranges of 14 parameters were only doubled. From the results of
a first round of GA runs with the factor-4-enlarged ranges and from runs of 
the ``original'' ranges from the van Duin group, 
new ranges were obtained by using the minimum and maximum values of
each parameter of the best individual obtained from the 10 runs. In the 
following, this strategy is referred to as ``fitting procedure''. This 
procedure was repeated for the factor-4-enlarged ranges.
The result can be seen in
Fig.~\ref{figure:range_fitting}. Obviously, a massive enlargement of
each parameter range {led} to significantly worse performance. For example, after 800 iterations, the runs with an enlargement
of factor 8 yielded individuals with error sums similar to
the \emph{randomly} created initial individuals of the runs with the ``original'' ranges
obtained from the van Duin group. However, the fitting procedure of the ranges was
quite successful. After two iterations of it, the ranges, which originally were
enlarged by factor of 4, {led} to runs which gave results comparable with the results
of runs with the original van Duin ranges. A fitting procedure of the latter runs
produced even better results. 

\section{Production procedure and results for SiOH}
\label{section:SiOH}

\subsection{Production procedure}
\label{subsection:SiOH.production}

According to the results of section~\ref{section:GAopt} and preliminary tests
prior to that, a rough strategy was evolved: At first, 12 production runs were
executed, 
with 4900 iterations without local optimization, followed by 100 iterations 
with local optimization using the \textsc{bobyqa}-routine and a loose 
convergence criterion. The iteration number was larger than in the 
procedures in section~\ref{section:GAopt.tune}, to exploit the most of each run. As 
noted, the strong initial progress typical for quasi-random search algorithms
keeping track of the best-so-far solution leveled off at 
about 600 to 800 iterations. However, it is also typical that such algorithms
still show occasional progress in the later stages, which is worthwhile to
attain during production, unless the overall timing becomes excessive.
The population sizes were 30 and 100 (see 
the discussion in section~\ref{section:GAopt.tune}). The parameter ranges
were obtained by the ``fitting procedure'' of 
the test runs with the original ranges, see section~\ref{section:GAopt.ranges}. 
Afterwards, the best individuals of each run (about 400) were selected for an 
additional optimization with the \textsc{newuoa}-routine and a tighter 
convergence criterion. As mentioned in section~\ref{section:methods.local}, the (more 
compute-intensive) \textsc{newuoa}-routine has no constraints so that 
parameters with 
values near the artificial parameter bounds can ``relax'' beyond those
bounds. Beginning with randomly  
created individuals with error sums between $2\cdot10^3$ and $1\cdot10^7$, the 
optimization ended after this phase at a best error sum of 4268. 

Then, a new starting population was selected from these optimized individuals, 
according to criteria of small error sum and large diversity, i.e.\ large 
{Euclidean distances ($L^2$ distance)} between one individual vector and all other vectors.
The ranges of each parameter were fitted to 
this starting population. New runs were executed with two selected populations 
with a size of 30 and 122, respectively, and the procedure mentioned above was 
repeated. Although the starting population was specified and not randomly 
created in this phase, the optimization runs showed quite unequal error sums 
for the best individual, which stresses the indeterministic character of global 
optimization. During these optimizations, we experienced a \emph{higher} error 
sum after local optimization with the \textsc{bobya}, the \textsc{soppe} and 
the older Powell routine (cf.\ section \ref{section:methods.local}). This should not
happen in a minimization routine but (without explicit tests) it cannot be
strictly avoided when finite-size steps are taken, and in particular if no
analytical gradient information is available. Since we observed such behavior
only for vectors with small error sums, we conclude{d} that this is also a
symptom of being close to the lower limit of global optimization.
Only the \textsc{newuoa}-routine showed small optimization success.

Additionally, the runs needed more iterations for a substantial decrease of the 
error sum --- iteration numbers between 15,000 and 30,000 were common. However, 
also the needed time for each iteration was diminished drastically, because all 
individuals in the population already had good error sums. Therefore, 
geometry optimizations that {were} part of the error sum calculation
needed less iterations because the minimum geometry with the ReaxFF-potential 
differed not so much from the starting-point geometry obtained by quantum 
mechanical calculation. An optimization run with the 
random populations took about seven days, whereas a run with the
starting population selected as explained and with 10 times more iterations with local 
optimization only took about three days. It was also observed that at later
stages of these runs a lot of children with very large error sums were created by 
crossover or mutation. Therefore, the program was slightly adjusted so that only 
children with sufficiently small error sum were locally optimized. Additionally, 
only the best 30\% of the population were locally optimized during the initialization 
of the local optimization phase, to accelerate the local optimization 
phase.

After three iterations of this  procedure with an increasing GA-iteration 
number 
and a decreasing convergence criterion for the local optimization routines, an 
individual with an error sum of 3196 was finally obtained. 

{Since progress to even better results appeared to become
increasingly difficult, we decided 
to test if this error sum could be further improved upon
by allowing more parameters than the originally pre-set 67 to be optimized.} 
Based on the 
purpose of each force field parameter, 124 additional parameters were selected 
that could have an influence on the 
SiOH force field. Therefore, 191 parameters now needed to be optimized. The 
first step to achieve this was to set up ranges for the extra parameters. 
To simulate a realistic situation where knowledge about parameter ranges may
be scarce or absent, we resorted to
comparing values of these 
parameters with values of similar parameters or of other force fields, and 
arranged ranges based on that. However, randomly created individuals with 
parameter values within these ranges led to force fields with unphysical 
behavior and program crashes. After further manual adjustment, the ranges 
mostly yielded ``stable'' individuals but with very high error sums. 
Therefore, 30 GA runs with 2000 iterations in each case were executed and the 
best 20 individuals with error sums at about $3\cdot 10^5$ were used to fit the 
parameter ranges for the first time. Then, two production rounds were
performed, following the aforementioned procedure, with iterations between
5000 and 10,000.  For the optimization runs with 67 parameters to be 
optimized, all values of other parameters were fixed and therefore the same 
for all individuals. Hence, only one (the best) vector of the outcome of the 
optimization process with 67 parameters to be optimized was used and added to 
the starting population in the second production round. Otherwise, there would 
have been a change in the values of the 124 extra parameters to be optimized 
only due to mutation, but not to crossover, which is the main operator of the 
GA. 

During this production, the values of the ranges for each parameter were 
successively inserted into some individuals from the 
global optimization, and the error sum was calculated, 
to check whether the parameter has an influence on the error sum or not. 34 of 
the additional parameters showed no large influence and were therefore 
excluded from further optimization. 
Otherwise, these parameters could have arrived at
not controllable values within their ranges. This would 
have had no influence on the behavior of the molecules tested in the training 
set but it could potentially affect the force-field performance for other
molecules outside the training set. 

After local optimization, an individual with the smallest error sum of 2807 was 
obtained. By doing further optimizations, this error sum could have been
improved further. However, this procedure was only meant as a demonstration of what
can be 
achieved when focusing on a single system. The individuals that were obtained by 
an optimization of 
the extended set of 191 parameters cannot be used outside of the SiOH-force field, since other force 
fields also use those additional parameters but with different values (cf.\ 
section \ref{section:intro}).

\subsection{Results}
\label{subsection:SiOH.results}

\begin{figure}[htbp]
  \centering
  \includegraphics{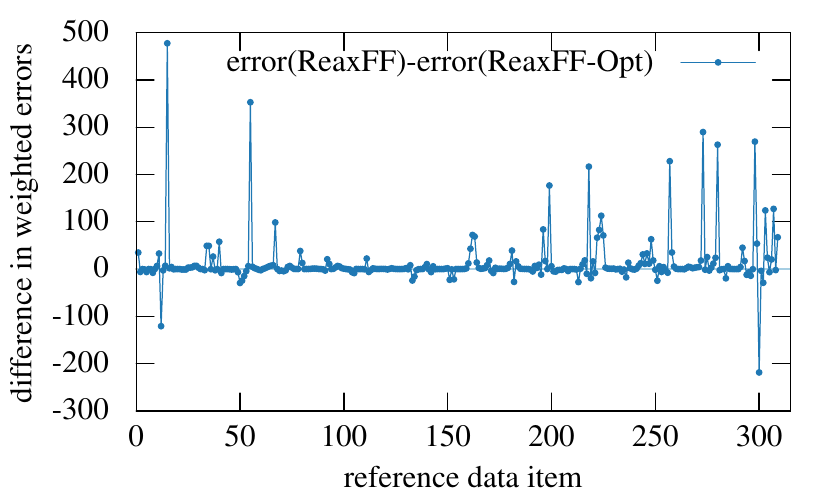}
  \begin{minipage}{0.8\textwidth}
  \caption{\label{figure:overview}Differences in individual error sum
    contributions (eq.~\ref{eq:error_sum}) between the original SiOH ReaxFF
    force field\cite{SiOH2} (``ReaxFF'', error sum of 6455) and our
    newly GA-fitted ReaxFF force field (``ReaxFF-Opt'', error sum of 2807),
    as a function of the (arbitrary) ordinal number of the entries in the
    training set.}
  \end{minipage}
\end{figure}

In Fig.~\ref{figure:overview}, a more fine-grained view on the results is
given, in terms of the individual weighted entries comprising the error sum of
eq.~\ref{eq:error_sum}, shown as the difference between the error items of the 
originally published
SiOH force field \cite{SiOH2} and the corresponding ones of our newly
GA-fitted one. Most of these error differences are close to zero, i.e., our
new results do not constitute a dramatic improvement. This is to be expected;
it re-emphasizes that the production of the original SiOH force field was done
with care and already provides a very good fit to the reference data, leaving
little room for further improvement, within the possibilities of the ReaxFF
functional form. Nevertheless, there are about two dozen reference data items
where there are substantial differences in the weighted errors. For only two
of them (no.~12 and 300), the graph in Fig.~\ref{figure:overview} peaks into
the negative direction, meaning that the error in our new fit is larger than
it was before. In all other cases (positive peaks), the error contributions
have become smaller. In addition to the overall error sum values, this is
further support for our claim that this new fit actually is better.
To properly judge this outcome, it should be recalled that the
originally published force-field already was the well-converged final result
of a diligent, compute-intensive project, performed by experienced researchers
in the group of the ReaxFF inventor. The GA-fitting in the Hartke group,
however, started with no experience in ReaxFF force-field fitting, and only
took about two months real-time within a BSc thesis project 
\cite{henrik},
including all the strategy-exploration work described partially in section
\ref{section:GAopt}. If the calculations necessary for the complete GA-fitting
were re-done by us now, they would take about 2 weeks on 200
processors.

Further insights into the new GA-fitted results become accessible in still more
fine-grained data. To this end, we show a semi-random selection of potential
energy surface cuts along certain coordinates, for certain molecular systems
from the training set. Obviously, in some cases, our GA fit
leads to substantial improvements, even of a qualitative nature. In others,
only smaller quantitative improvements are visible or even no improvement at
all, despite qualitative defects compared to the reference data.

\begin{figure}[htbp]
  \centering
  \includegraphics[width=0.8\textwidth]{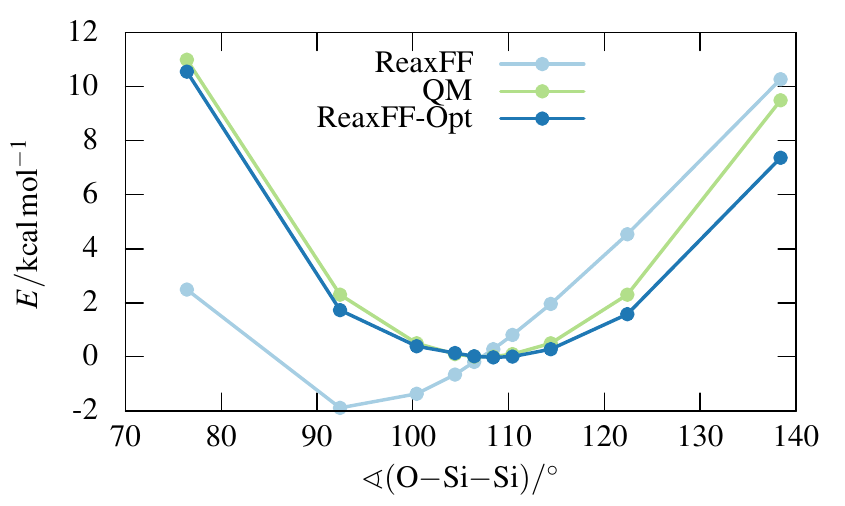}
  \begin{minipage}{0.8\textwidth}
  \caption{\label{figure:O-Si-Si_angle}Potential energy as a function of
    O-Si-Si bond angle in \ce{HO-SiH2-SiH3}, for the reference data
    (B3LYP/6-31G$^\ast$) (green), the original SiOH ReaxFF data\cite{SiOH2}
    (light blue, error sum of 6455), and the newly GA-fitted ReaxFF data (dark 
  blue, error sum of  2807).}
  \end{minipage}
\end{figure}

\begin{figure}[htbp]
  \centering
  \includegraphics[width=0.8\textwidth]{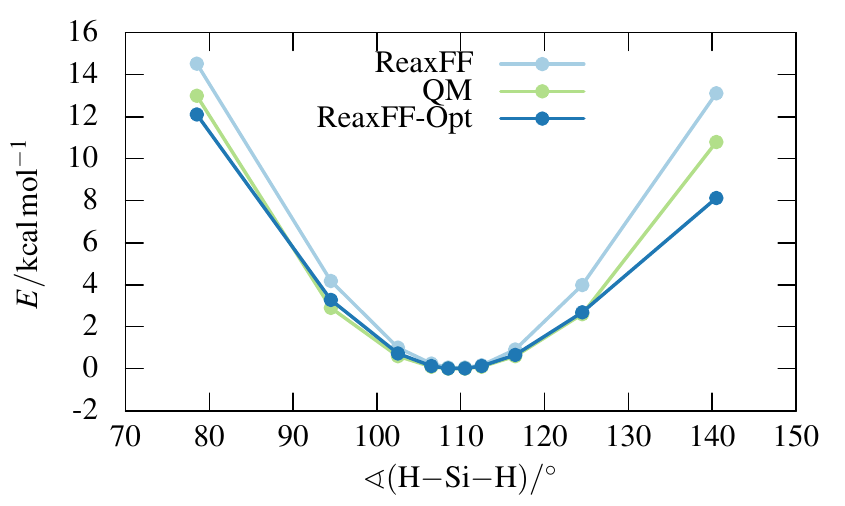}
  \begin{minipage}{0.8\textwidth}
  \caption{\label{figure:H-Si-H_angle}As Fig.~\ref{figure:O-Si-Si_angle},
    for the H-Si-H bond angle in \ce{SiH4}.}
  \end{minipage}
\end{figure}

\begin{figure}[htbp]
  \centering
  \includegraphics[width=0.8\textwidth]{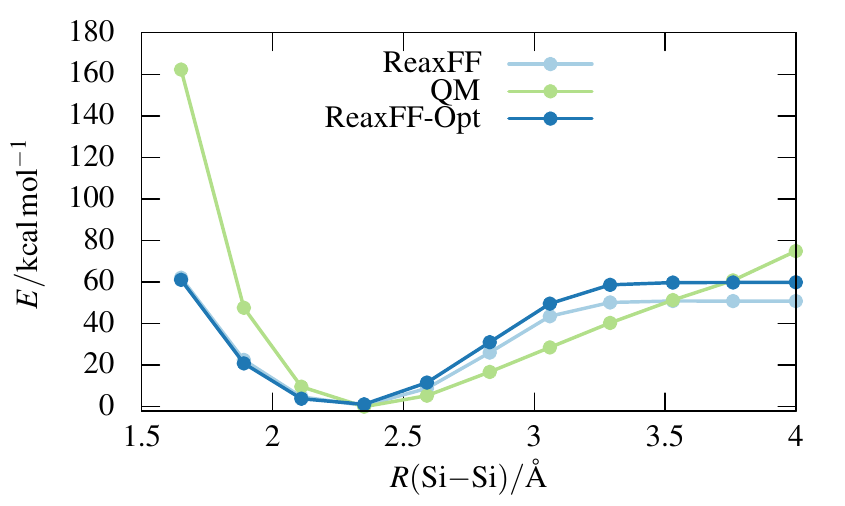}
  \begin{minipage}{0.8\textwidth}
  \caption{\label{figure:Si-Si_bond}As Fig.~\ref{figure:O-Si-Si_angle},
    for the Si-Si bond distance in \ce{H3Si-SiH3}.}
  \end{minipage}
\end{figure}

\begin{figure}[htbp]
  \centering
  \includegraphics[width=0.8\textwidth]{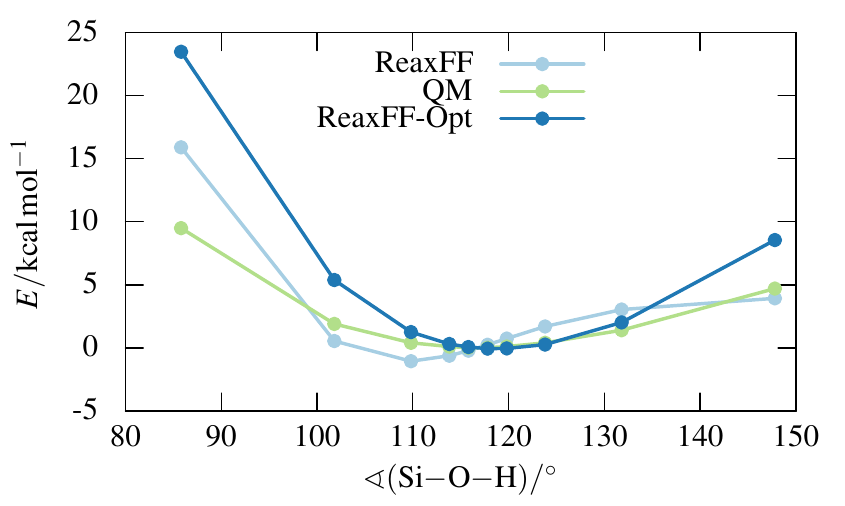}
  \begin{minipage}{0.8\textwidth}
  \caption{\label{figure:Si-O-H_angle}As Fig.~\ref{figure:O-Si-Si_angle},
    for the Si-O-H bond angle in \ce{H3Si-O-H}.}
  \end{minipage}
\end{figure}

We show four examples of this kind in Figs.~\ref{figure:O-Si-Si_angle} through
\ref{figure:Si-O-H_angle}. In Fig.~\ref{figure:O-Si-Si_angle}, our newly
optimized ReaxFF data obviously constitute a substantial improvement over the
previously published ones \cite{SiOH2}, not just quantitatively but even
qualitatively. In Fig.~\ref{figure:H-Si-H_angle}, the diagnosis is less clear,
but upon closer inspection only the rightmost data point prevents us from
reaching the same conclusion again. In
Fig.~\ref{figure:Si-Si_bond}, both
ReaxFF curves manage to get the location of the minimum correctly, but fail
qualitatively in other regions; our new fit is no improvement here. In
Fig.~\ref{figure:Si-O-H_angle}, the overall shape of our new ReaxFF curve is
better but the data points of the old force field stayed closer to the
reference data overall.

{
For interested readers, we provide our two best force fields (with error sums
of 2807 and 3196, respectively) and full information on the training set
(geometries and properties) and on the parameters to be optimized (including
their variation ranges) in the supplementary information.}

\clearpage

\section{Conclusions}
\label{section:conclusions}

\subsection{{Summary}}
\label{section:conclusions.summary}

We have presented an unbiased, global GA optimization of a ReaxFF
force field for SiOH species. A set of 304 reference structures (containing a
mixture of species, ranging from simple silanes to quartz) and a set of 309
reference properties for them (a mixture of geometry data, partial charges,
crystal cell parameters, and relative energies) were taken as given, as well as
a realistic set of 67 force-field parameters to be optimized, within given
ranges. 

Beyond these data, no further prior information was utilized in the GA
search. Nevertheless, our best GA-optimized force fields constitute
further improvements upon the (already good) SiOH-ReaxFF force fields
published previously. { For the force field from
Ref.~\cite{SiOH2}, the error sum was 6455, but later re-optimizations using
\textsc{soppe}, prompted by our GA work, could bring this number down to just
under 4000. Our best GA results range between 2800 and 3200. From the
behavior observed in our optimizations, we assume that
no significant further improvement will be possible, with the selection of
parameters and reference data items set by the van Duin group. Error sum
values of 2800 (and possibly lower) are accessible with an 
extension of the set of parameters to be optimized, to a total of 191, but this
compromises the compatibility of the resulting SiOH force field with other
ReaxFF parametrizations, and hence was not pursued further.} 

{
Since no expert knowledge about the force field in general or the purpose of
its parameters in particular is necessary for our GA parameter optimization
approach, it} provides easier access to 
force-field optimization for less experienced users, and it also helps
experienced force-field optimizers to arrive at better solutions, closer to
global minima. {As one step towards end-user distribution of our
code, integration of it into the ADF package \cite{ADF_ReaxFF} currently is
underway.}

In GA applications of this kind, detailed tuning of the GA is less important
than providing suitable ranges in which the force-field parameters are allowed
to vary. Nevertheless, as demonstrated here, it is also possible and realistic
to iteratively improve upon badly guessed initial parameter ranges:
{
A first
round of GA runs is started with the initially guessed variation ranges (which
can be considerably too wide)
 and with random initialization of the population within
these ranges. From the resulting final populations, the best individuals are
selected. Their parameter value variations allow for a realistic estimate of
narrower variation ranges. These best individuals constitute the starting
population for a second round of GA runs, employing the narrower variation
ranges, which improves GA efficiency is considerably.
After a few iterations, the variation ranges have shrunk to intervals that
allow for 
arriving at results of the same quality as with more suitable initial
parameter ranges.}

\subsection{{Outlook}}
\label{section:conclusions.outlook}

{
The Gly test case mentioned briefly in section \ref{section:methods.testcases}
is also based upon recent work in the van Duin group
\cite{Gly}, where a ReaxFF parametrization for the amino acid glycin was
developed, in order to investigate the water-mediated 
neutral $\rightleftharpoons$ zwitterionic tautomerization in aqueous
solution. Here, the
reference data set consists of 2049 items, and 299 parameters are to be
varied, producing a very much larger search space than for the other two
cases. Work on the Gly test case is in progress. From its current status, we
cannot yet decide whether the same techniques demonstrated here for the SiOH
test case also allow us to finish the Gly test case successfully or if such a
large search space
would need impracticably large amounts of computer time. 
One possible remedy in
the latter situation may be iterative cycling over smaller-sized subproblems,
which was shown to be successful in cluster structure optimization
\cite{Michaelian}.}

As noted in section \ref{section:methods.parallel}, for our current master-slave
program version, parallel speedup is worthwhile for a few dozen slave
processes. In future developments of our code, we plan to change
this by handing out chunks of several reference data items. This will improve
the ratio of computation to communication and hence will allow for better
scalability. This will be improved upon further by also implementing the
(simpler) parallelism at the level of GA individuals. Parallelism at the
lowest level, at the calculation of ReaxFF energies, will be included only
later, since it is most worthwhile only for large systems, which do not tend
to be part of reference data sets for parameter fitting.

For each individual and independent of parallelization issues, further speedup
can be realized by calculating the error sum ``on the fly'', as each reference
data item is returned to the master process. Currently, this is done only
after every item has been processed. The on-the-fly option would make it
possible to refrain from further (probably costly) evaluations of reference
data items if the error sum already is too large to be competitive. Of course,
this destroys some information on the shape of the cost function, but only in
its worst regions. It can be argued that the guidance offered by these regions
is limited to non-existent anyway, in a scheme like a GA where the best
schemata tend to be strongly favored. No information of any consequence is
lost, however, if the on-the-fly error sum cutoff is placed at the bad end of
the ordered pool. {In test calculations for the SiOH case,
  performed after the studies described in this article, this has incurred
  substantial savings in computer time. It remains to be seen if this can be
  generalized.}

\subsection*{Acknowledgements}

BH thanks Alexei Yakovlev (SCM company, Netherlands) for his patient and
competent help with incorporating our techniques into the ADF program package.

\newpage

\section*{Graphical abstract}

\begin{figure}[htbp]
  \centering
  \includegraphics[width=0.4\textwidth]{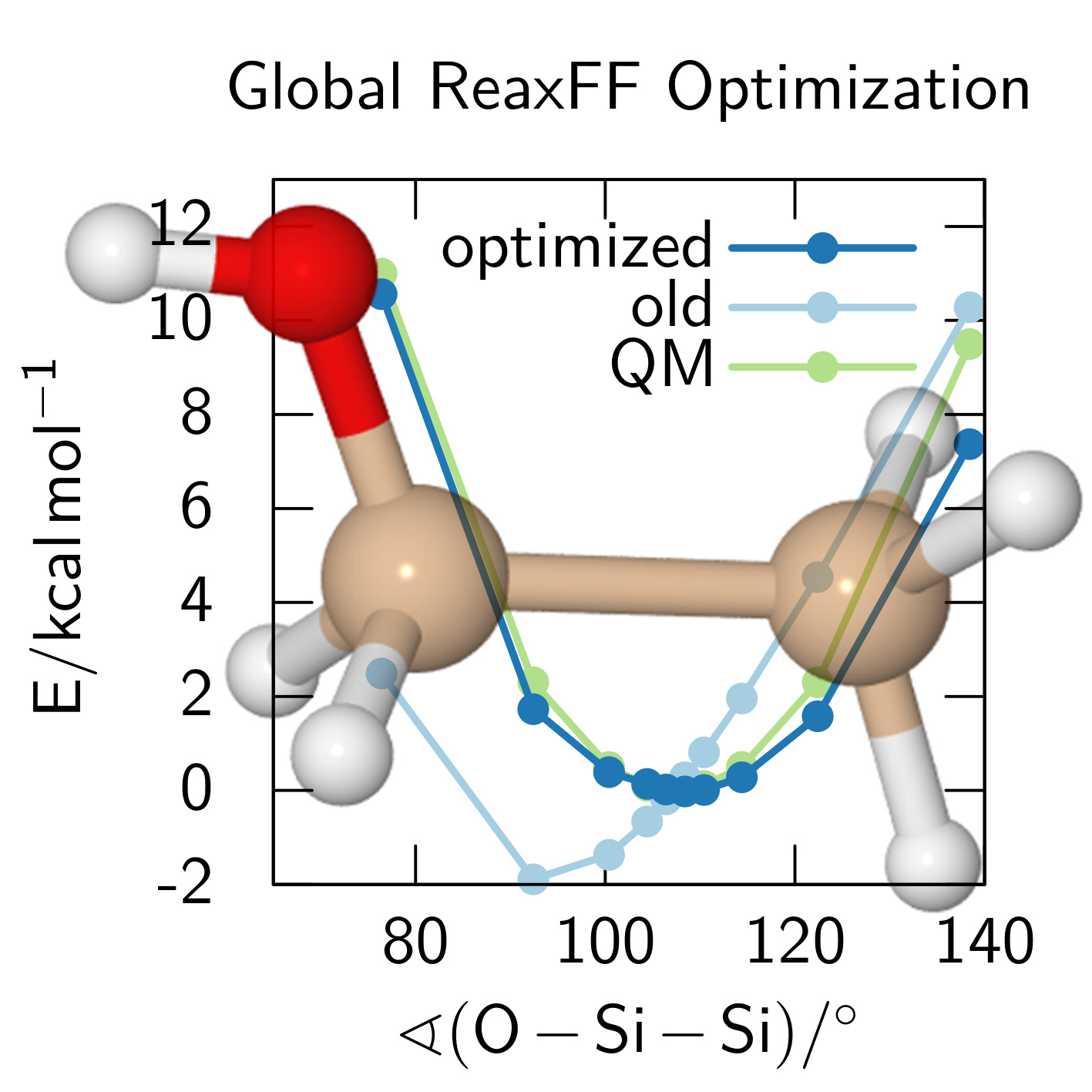}
\end{figure}

Chemical reactions can be described using reactive force fields. The challenge
of fitting the reactive force field parameters to reference data has been
addressed with global optimization using genetic algorithms. Without expert
knowledge to guide the search, this results in superior agreement with the
reference data. This contribution demonstrates this approach for molecules
containing silicon, oxygen and hydrogen atoms, employing ReaxFF.

\end{document}